\documentclass[12pt,a4paper]{article}
\usepackage{jheppub}
\usepackage{amssymb}
\usepackage{amsmath}
\usepackage{topcapt}
\usepackage{bbm}

\DeclareSymbolFont{AMSb}{U}{msb}{m}{n}
\DeclareMathSymbol{\IN}{\mathbin}{AMSb}{"4E}
\DeclareMathSymbol{\IZ}{\mathbin}{AMSb}{"5A}
\DeclareMathSymbol{\IR}{\mathbin}{AMSb}{"52}
\DeclareMathSymbol{\Q}{\mathbin}{AMSb}{"51}
\DeclareMathSymbol{\II}{\mathbin}{AMSb}{"49}
\DeclareMathSymbol{\IC}{\mathbin}{AMSb}{"43}
\DeclareMathSymbol{\IP}{\mathbin}{AMSb}{"50}
\DeclareMathSymbol{\IH}{\mathbin}{AMSb}{"48}
\DeclareMathSymbol\IA{\mathalpha}{AMSb}{"41}
\DeclareMathSymbol\IS{\mathalpha}{AMSb}{"53}

\newcommand{\thetp}{\theta_{{\scalebox{0.65}[.65]{\(+\)}}}}
\newcommand{\thetm}{\theta_{{\scalebox{0.65}[.65]{\(-\)}}}}
\newcommand{\phip}{\phi_{{\scalebox{0.65}[.65]{\(+\)}}}}
\newcommand{\phim}{\phi_{{\scalebox{0.65}[.65]{\(-\)}}}}
\def\N{{\cal N}}

\title{Flavours in global Klebanov--Witten background}

\author[a]{Dimo Arnaudov,} 
\author[b]{Veselin Filev,}
\author[a,c]{Radoslav Rashkov,}
\affiliation[a]{Department of Physics, Sofia University, 5 J. Bourchier Blvd, 1164 Sofia, Bulgaria}
\affiliation[b]{School of Theoretical Physics, Dublin Institute for Advanced Studies, 10 Burlington Rd, 4 Dublin, Ireland}
\affiliation[c]{Institute for Theoretical Physics, Vienna University of Technology, Wiedner Hauptstr. 8-10, 1040 Vienna, Austria}

\emailAdd{dlarnaudov@phys.uni-sofia.bg}
 \emailAdd{vfilev@stp.dias.ie}
\emailAdd{rash@hep.itp.tuwien.ac.at}

\date{\today}

\abstract{
We probe flavour branes in the Klebanov--Witten background in global coordinates. We show that supersymmetric probes correspond to massless flavours and analyse their spectrum of fluctuations. Our studies suggest that the mesons are dissociated by the Casimir energy. The spectrum is equidistant, with a ground state given by the conformal dimension of the operator dual to the fluctuations. We show that the equidistant structure of the spectrum arises from the addition of the angular momenta of the fundamental fields composing the meson operators.}

\keywords{AdS/CFT correspondence, Gauge/gravity correspondence}

\begin{document}
\maketitle

\section{Introduction}
The initial form of the AdS/CFT correspondence concerns the near-horizon limit of $N$ D3--branes, which entails the study of dynamics in $AdS_5\times S^5$ background. In ref.~\cite{Karch:2002sh} probe D7--branes were added to this geometry with the effect of introducing hypermultiplets in the fundamental representation of the gauge group on the dual side. The condition for finiteness of the fluctuations of probe branes leads to the spectra of holographic mesons \cite{Kruczenski:2003be}. Subsequent generalizations were given in \cite{{Kirsch:2004km},{Arean:2006pk},{Myers:2006qr},{Hoyos:2006gb}}.

Another fruitful possibility is the embedding of D5--brane probes, which wrap an $AdS_4\times S^2$ part of $AdS_5\times S^5$ \cite{DeWolfe:2001pq}. The dual gauge theory at hand is a superconformal ``defect'' theory \cite{Erdmenger:2002ex} with corresponding hypermultiplets confined to a (2+1)-dimensional subspace.

In the present paper we embed probe D7-- and D5--branes in global $AdS_5\times T^{1,1}$ background. On the dual side we have an $\N=1$ supersymmetric quiver gauge theory, the Klebanov--Witten model, living on a three-sphere \cite{Klebanov:1998hh}. Holographic studies of flavoured gauge theories on compact manifolds were initiated in refs.~\cite{{Karch:2009ph},{Karch:2006bv}}, where the most supersymmetric case of flavoured ${\cal N}=4$ SYM theory on a three-sphere\footnote{Note that the flavours break the ${\cal N}=4$ supersymmetry down to ${\cal N}=2$ one. However, since the term ${\cal N}=4$ SYM uniquely specifies the field content of the adjoint fields, throughout the paper we will refer to this ${\cal N}=2$ supersymmetric theory as the flavoured ${\cal N}=4$ theory, keeping in mind that half of the supersymmetry is broken.}  was analysed. A detailed study of the meson spectrum was performed in ref. \cite{Erdmenger:2010zm}, where it was shown that for the chiral primaries, the lowest level is given by the scaling dimension of the operator corresponding to the fluctuations. It was also shown that the spectrum exhibits an equidistant structure resulting from the summation of the angular momenta of the fundamental fields composing the mesons.\footnote{See also refs. \cite{{PremKumar:2011ag},{Chunlen:2012zy},{Filev:2012ch}} for thermodynamic studies with external control parameters.}  In this paper we perform an analogous study for the flavoured Klebanov--Witten model on a three-sphere.\footnote{For studies of the flavoured Klebanov--Witten model in the Poincare patch, relevant to our studies, see refs.~\cite{Apreda:2006bu}-\cite{Filev:2013vka}. } Our study is particularly interesting because of the significantly less amount of supersymmetry (${\cal N}=1/2$) of the flavoured  Klebanov--Witten model. Interestingly, the meson spectra exhibit the same properties as for the flavoured ${\cal N}=4$ theory, the only difference being that the conformal dimensions of the fields composing the meson states cannot be determined from the free field theory on an $S^3$, which is natural because of the absence of non-renormalisation theorems for less than two supersymmetries.

This paper is organised as follows. In section 2 we investigate which probe brane embeddings are supersymmetric via kappa-symmetry techniques. In section 3 we calculate analytically the meson spectra at zero quark mass for probe D7--brane fluctuations on the gravity side. Analogously, in section 4 we obtain the scalar bosonic spectrum for D5--brane fluctuations. We conclude with a brief discussion on the results in section 5. Finally, there is an appendix, in which we give details about the parametrization of $T^{1,1}$ and the dual gauge theory.

\section{Supersymmetric probes}
In this section we analyse the kappa symmetry of the probe branes. Our analysis follows closely the one performed in \cite{Arean:2004mm}, where the Killing spinors in both global and Poincare coordinates are discussed.

\subsection{Killing spinor.}
The metric of the Klebanov--Witten geometry is given by:
\begin{equation}
ds^2=R^2(-\cosh^2\rho dt^2+d\rho^2+\sinh^2\rho d\Omega_3^2)+R^2ds_{T^{1,1}}^2\ ,
\end{equation}
where $ds_{T^{1,1}}^2$ is the metric on $T^{1,1}$ given by:
\begin{equation}
ds_{T^{1,1}}^2=\frac{1}{6}\sum\limits_{i=1}^2\left(d\theta_i^2+\sin^2\theta_i\,d\phi_i^2\right)
+\frac{1}{9}\!\left(d\psi+\sum\limits_{i=1}^2\cos\theta_i\,d\phi_i\right)^2. \label{T11}
\end{equation}
Here the ranges of the angles are: $0<\theta_i<\pi$, $0<\phi_i<2\pi$ and $0<\psi<4\pi$. Using the one-forms in \cite{Arean:2004mm}:
\begin{equation}
\sigma^1=d\theta_1\ ,\qquad\sigma^2=\sin\theta_1d\phi_1\ ,\qquad\sigma^3=\cos\theta_1d\phi_1\ ,
\end{equation}
associated to a two-sphere and the forms
\begin{eqnarray}\nonumber
w^1&=&\sin\psi\sin\theta_2\,d\phi_2+\cos\psi\,d\theta_2\ ,\\
w^2&=&-\cos\psi\sin\theta_2\,d\phi_2+\sin\psi\,d\theta_2\ ,\\ \nonumber
w^3&=&d\psi+\cos\theta_2\,d\phi_2\ ,
\end{eqnarray}
associated to a three-sphere, one can write down the metric of $T^{1,1}$ in the following form:
\begin{equation}
ds_{T^{1,1}}^2=\frac{1}{6}\left((\sigma^1)^2+(\sigma^2)^2+(w^1)^2+(w^2)^2\right)+\frac{1}{9}\left(\sigma^3+w^3\right)^2\,.
\end{equation}
If we parameterize the three-sphere of the global AdS$_5$ as
\begin{equation}
d\Omega_3^2=d\alpha_1^2+\sin^2\alpha_1\!\left(d\alpha_2^2+\sin^2\alpha_2\,d\alpha_3^2\right)\,,
\end{equation}
we can choose the following tetrads \cite{Arean:2004mm}:
\begin{eqnarray}
\label{frame}
&e^t&=R\cosh\rho\,dt\ ,~~~~e^\rho=Rd\rho\ ,\\ \nonumber
&e^{\alpha_1}&=R\sinh\rho\,d\alpha_1\ ,\\ \nonumber
&e^{\alpha_2}&=R\sinh\rho\sin\alpha_1d\alpha_2\ ,\\ \nonumber
&e^{\alpha_3}&=R\sinh\rho\sin\alpha_1\sin\alpha_2\,d\alpha_3\ ,\\ \nonumber
&e^i&=\frac{R}{\sqrt{6}}\,\sigma^i\,,~~~~e^{\hat i}=\frac{R}{\sqrt{6}}\,w^i\,,~~~~i=1,2\ ,\\ \nonumber
&e^3&=\frac{R}{3}\left(\sigma^3+w^3\right)\,.
\end{eqnarray}
Let us define the matrix:
\begin{equation}
\gamma_*=\Gamma_t\Gamma_{\rho}\Gamma_{\alpha_1\,\alpha_2\,\alpha_3}\ ,
\end{equation}
where the $\Gamma$'s are the flat gamma matrices associated to the choice of frame (\ref{frame}). One can write down the Killing spinor of the background as:
\begin{equation}
\label{kilspin}
\varepsilon=e^{-\frac{i}{2}\rho\Gamma_{\rho}\gamma_*}e^{-\frac{i}{2}t\Gamma_t\gamma_*}e^{-\frac{\alpha_1}{2}\Gamma_{\alpha_1\rho}}
e^{-\frac{\alpha_2}{2}\Gamma_{\alpha_2\alpha_1}}e^{-\frac{\alpha_3}{2}\Gamma_{\alpha_3\alpha_2}}\eta\ ,
\end{equation}
where $\eta$ is a constant spinor satisfying the following conditions:
\begin{equation}
\label{projection}
\Gamma_{12}\,\eta=i\eta\ ,~~~~\Gamma_{{\hat 1}{\hat 2}}\,\eta=-i\eta\ .
\end{equation}
Notice that $\Gamma_{12}$ and $\Gamma_{\hat 1\hat 2}$ commute with the matrices on the right-hand side of equation (\ref{kilspin}) and therefore the Killing spinor also satisfies the conditions (\ref{projection}):
\begin{equation}
\label{projectioneps}
\Gamma_{12}\,\varepsilon=i\varepsilon\ ,~~~~\Gamma_{{\hat 1}{\hat 2}}\,\varepsilon=-i\varepsilon\ .
\end{equation}

\subsection{Kappa-symmetry matrix.}
The condition for a probe brane embedding to preserve some fraction of the supersymmetry of the background is:
\begin{equation}
\label{projeps}
\Gamma_{\kappa}\,\varepsilon=\varepsilon\ ,
\end{equation}
where $\Gamma_{\kappa}$ is given by:
\begin{equation}
\label{gammak}
\Gamma_{\kappa}=\frac{1}{(P+1)!\sqrt{-g}}\,\varepsilon^{\mu_1\dots\mu_{p+1}}(\tau_3)^{\frac{p-3}{2}}i\tau_2\otimes\gamma_{\mu_1\dots\mu_{p+1}}\ .
\end{equation}
In equation (\ref{gammak}) $g$ is the determinant of the induced metric $g_{\mu\nu}$, given by:
\begin{equation}
g_{\mu\nu}=\partial_{\mu}X^M\partial_{\nu}X^N\,G_{MN}\ ,
\end{equation}
and $\gamma_{\mu_1\dots\mu_{p+1}}$ is an antisymmetrised product of the induced gamma matrices $\gamma_{\mu}$, given by:
\begin{equation}
\gamma_{\mu}=\partial_{\mu}X^{M}E^{\bar N}_{M}\Gamma_{\bar N}\ ,
\end{equation}
where $E^{\bar N}_M$ are the components of the one-form frame given in equation (\ref{frame}). The $\tau_a$ matrices in equation (\ref{gammak}) are Pauli matrices acting on the vector $\left(\begin{array}{c}\varepsilon_1\\ \varepsilon_2\end{array}\right)$, where $\varepsilon_1$ and $\varepsilon_2$ are the real and imaginary parts of the complex spinor $\varepsilon$.

\subsection{Kappa symmetry for a D5--brane probe.}
For a D5--brane probe the kappa-symmetry matrix (\ref{gammak}) takes the form:
\begin{equation}
\Gamma_{\kappa}=\frac{1}{6!\sqrt{-g}}\,\varepsilon^{\mu_1\dots\mu_6}\tau_1\otimes\gamma_{\mu_1\dots\mu_6}\ .
\end{equation}
\\
Note that $\tau_1\left(\begin{array}{c}\varepsilon_1\\ \varepsilon_2\end{array}\right)=\left(\begin{array}{c}\varepsilon_2\\ \varepsilon_1\end{array}\right)$, which are the components of the complex spinor $i\varepsilon^*$. Therefore we can consider the action of $\Gamma_{\kappa}$ in complex notation \cite{Arean:2004mm}:
\begin{equation}
\label{GammakD5}
\Gamma_{\kappa}\,\varepsilon=\frac{i}{6!\sqrt{-g}}\,\varepsilon^{\mu_1\dots\mu_6}\gamma_{\mu_1\dots\mu_6}\varepsilon^*\,.
\end{equation}
We are interested in space-filling D5--brane embeddings. Furthermore, the analysis of the supersymmetries of the D5--brane embeddings in the flat case studied in ref. \cite{Arean:2004mm} shows that supersymmetric embeddings give rise to co-dimension one defect field theory. In global coordinates this corresponds to a D5--brane embedding wrapping a maximal two-sphere of the AdS$_5$ three-sphere. Therefore we consider the following ansatz for the D5--brane embedding:
\begin{equation}
\xi^{\mu}=(t,\rho,\alpha_1,\alpha_3,\theta_1,\phi_1)\ ,
\end{equation}
and
\begin{equation}\label{ansatzD5}
\theta_2=\theta_2(\theta_1,\phi_1)\ ,~~~\phi_2=\phi_2(\theta_1,\phi_1)\ ,~~~\psi=\rm{const}\ ,~~~\alpha_2=\frac{\pi}{2}\ .
\end{equation}
For this ansatz the kappa-symmetry matrix (\ref{GammakD5}) becomes:
\begin{equation}
\Gamma_{\kappa}\,\varepsilon=\frac{i}{\sqrt{g_{X^2}}}\,\tilde\gamma_*\gamma_{\theta_1\,\phi_1}\varepsilon^*\,,
\end{equation}
where
\begin{equation}
\tilde\gamma_*=\Gamma_t\Gamma_{\rho}\Gamma_{\alpha_1}\Gamma_{\alpha_3}\ ,
\end{equation}
and $g_{X^2}$ is the determinant of the induced metric on the internal manifold $X^2\in T^{1,1}$ wrapped by the D5--brane. The matrix $\gamma_{\theta_1\phi_1}$ is the antisymmetrised product of $\gamma_{\theta_1}$ and $\gamma_{\phi_1}$ given by\footnote{Note that the expressions in equation (\ref{gamath1ph1}) are the same as in the flat case analysed in \cite{Arean:2004mm}.}:
\begin{eqnarray}
\nonumber
\gamma_{\theta_1}&=&\frac{L}{\sqrt{6}}\,\Gamma_1+\frac{L}{\sqrt{6}}\left(\cos\psi\partial_{\theta_1}\theta_2+\sin\psi\sin\theta_2\partial_{\theta_1}\phi_1\right)
\Gamma_{\hat 1}+\frac{L}{3}\cos\theta_2\partial_{\theta_1}\phi_2\,\Gamma_3+\\ \label{gamath1ph1}
&+&\frac{L}{\sqrt{6}}\left(\sin\psi\partial_{\theta_1}\theta_2-\cos\psi\sin\theta_2\partial_{\theta_1}\phi_2\right)\Gamma_{\hat 2}\ ,\\ \nonumber
\gamma_{\phi_1}&=&\frac{L}{\sqrt{6}}\sin\theta_1\Gamma_{2}+\frac{L}{\sqrt{6}}\left(\cos\psi\partial_{\phi_1}\theta_2
+\sin\psi\sin\theta_2\partial_{\phi_1}\phi_2\right)\Gamma_{\hat 1}+\\ \nonumber
&+&\frac{L}{3}\left(\cos\theta_1+\cos\theta_2\partial_{\phi_1}\phi_2\right)\Gamma_3
+\frac{L}{\sqrt{6}}\left(\sin\psi\partial_{\phi_1}\theta_2-\cos\psi\sin\theta_2\partial_{\phi_1}\phi_2\right)\Gamma_{\hat 2}\ .
\end{eqnarray}
One can check that $\gamma_{\theta_1\phi_1}$ commutes with all matrices acting on the right-hand side of equation (\ref{kilspin}), however $\tilde\gamma_*$ anticommutes with all matrices except the matrix $\Gamma_{\alpha_1\rho}$, with which it commutes. This will flip the sign in all of the exponents in equation (\ref{kilspin}) except the one involving $\Gamma_{\alpha_1\rho}$. On the other hand, in equation (\ref{GammakD5}) $\tilde\gamma_*$ is acting on the complex conjugate of the Killing spinor, which implies a careful choice of the basis for the gamma matrices. It turns out that choosing $\Gamma_t\,,\Gamma_{\rho}\,,\Gamma_{\alpha_1}\,,\Gamma_{\alpha_2}\,,\Gamma_{\alpha_3}\,,\Gamma_3$ to be imaginary and $\Gamma_1\,,\Gamma_2\,,\Gamma_{\hat 1}\,,\Gamma_{\hat 2}$ to be real is a viable choice, and the projection (\ref{projeps}) is equivalent to a projection by a constant operator on the constant spinor $\eta$. Let us briefly show this.

Substituting (\ref{kilspin}) and (\ref{GammakD5}) in the projection equation (\ref{projeps}) leads to:
\begin{eqnarray}
\label{projet1}
&&e^{-\frac{i}{2}\rho\Gamma_{\rho}\gamma_*}e^{-\frac{i}{2}t\Gamma_t\gamma_*}e^{-\frac{\alpha_1}{2}\Gamma_{\alpha_1\rho}}e^{\frac{\pi}{4}\Gamma_{\alpha_2\alpha_1}}
e^{\frac{\alpha_3}{2}\Gamma_{\alpha_3\alpha_2}}\frac{i\tilde\gamma_*\gamma_{\theta_1\phi_1}}{\sqrt{g_{X2}}}\eta^*=\\ \nonumber
&&=e^{-\frac{i}{2}\rho\Gamma_{\rho}\gamma_*}e^{-\frac{i}{2}t\Gamma_t\gamma_*}e^{-\frac{\alpha_1}{2}\Gamma_{\alpha_1\rho}}
e^{-\frac{\pi}{4}\Gamma_{\alpha_2\alpha_1}}e^{-\frac{\alpha_3}{2}\Gamma_{\alpha_3\alpha_2}}\eta\ ,
\end{eqnarray}
where we have used that $\Gamma_{\rho}\gamma_*\,,\Gamma_t\gamma_*\,,\Gamma_{\alpha_1\rho}\,,\Gamma_{\alpha_2\alpha_1}$ and $\Gamma_{\alpha_3\alpha_2} $ are real in the chosen basis. Using that
\begin{equation}
e^{\frac{\pi}{4}\Gamma_{\alpha_2\alpha_1}}e^{\frac{\alpha_3}{2}\Gamma_{\alpha_3\alpha_2}}=e^{-\frac{\pi}{4}\Gamma_{\alpha_2\alpha_1}}
e^{-\frac{\alpha_3}{2}\Gamma_{\alpha_3\alpha_2}}\Gamma_{\alpha_2\alpha_1}~\rm{and}~\Gamma_{\alpha_2\alpha_1}\tilde\gamma_*=\Gamma_{t\,\rho\,\alpha_2\,\alpha_3}\,,
\end{equation}
equation (\ref{projet1}) reduces to:
\begin{equation}
\label{proet}
\frac{i}{\sqrt{g_{X2}}}\,\Gamma_{t\,\rho\,\alpha_2\,\alpha_3}\gamma_{\theta_1\phi_1}\eta^*=\eta\ .
\end{equation}

Let us analyze the form of $\gamma_{\theta_1\phi_1}$. Using equations (\ref{gamath1ph1}) and the projections\footnote{Note that in our basis $\Gamma_{12}$ and $\Gamma_{\hat 1\hat 2}$ are real.}:
\begin{equation}
\label{projcomplx}
\Gamma_{12}\,\eta^*=-i\eta^*\,,~~~\Gamma_{\hat 1\hat 2}\,\eta^*=i\eta^*\,,
\end{equation}
one arrives at the following expression for $\gamma_{\theta_1\phi_1}\eta^*$:
\begin{equation}
\label{gammat1p1}
\gamma_{\theta_1\phi_1}\eta^*=-ic_1\,\eta^*+(c_2+ic_3)e^{i\psi}\Gamma_{1\hat 2}\,\eta^*+(c_4+ic_5)\Gamma_{13}\,\eta^*
+(c_6+ic_7)e^{i\psi}\Gamma_{\hat 1 3}\,\eta^*\,,
\end{equation}
where
\begin{eqnarray}
c_1&=&\frac{L^2}{6}\left[\sin\theta_1+\sin\theta_2(\partial_{\theta_1}\theta_2)(\partial_{\phi_1}\phi_2)
-\sin\theta_2(\partial_{\theta_1}\phi_2)(\partial_{\phi_1}\theta_2)\right]\,,\\ \nonumber
c_2&=&\frac{L^2}{6}\left[\sin\theta_1(\partial_{\theta_1}\theta_2)-\sin\theta_2(\partial_{\phi_1}\phi_2)\right]\,,~~~
c_3=-\frac{L^2}{6}\left[\partial_{\phi_1}\theta_2+\sin\theta_1\sin\theta_2(\partial_{\theta_1}\phi_2)\right]\,,\\ \nonumber
c_4&=&\frac{L^2}{3\sqrt{6}}\left[\cos\theta_1+\cos\theta_2(\partial_{\phi_1}\phi_2)\right]\,,~~~
c_5=\frac{L^2}{3\sqrt{6}}\sin\theta_1\cos\theta_2(\partial_{\theta_1}\phi_2)\ ,\\ \nonumber
c_6&=&\frac{L^2}{3\sqrt{6}}\left[\left(\cos\theta_1+\cos\theta_2(\partial_{\phi_1}\phi_2)\right)(\partial_{\theta_1}\theta_2)
-\cos\theta_2(\partial_{\phi_1}\theta_2)(\partial_{\theta_1}\phi_2)\right]\,,\\ \nonumber
c_7&=&-\frac{L^2}{3\sqrt{6}}\left[\left(\cos\theta_1+\cos\theta_2(\partial_{\phi_1}\phi_2)\right)\sin\theta_2(\partial_{\theta_1}\phi_2)
-\cos\theta_2\sin\theta_2(\partial_{\phi_1}\phi_2)(\partial_{\theta_1}\phi_2)\right]\,.
\end{eqnarray}
The main restriction on the form of $\gamma_{\theta_1\phi_1}$ comes from the requirement that the projections (\ref{proet}) and (\ref{projection}) commute. Given that $\Gamma_{12}$ and $\Gamma_{\hat 1 \hat 2}$ are real, this is possible only if the matrix acting on $\eta^*$ on the left-hand side of equation (\ref{proet}) anticommutes with both $\Gamma_{12}$ and $\Gamma_{\hat 1 \hat 2}\,$. Indeed, let us define:
\begin{equation}
\label{P}
{\cal P}\eta=A\,\eta^*\,,~~~A=\frac{i}{\sqrt{g_{X2}}}\,\Gamma_{t\,\rho\,\alpha_2\,\alpha_3}\gamma_{\theta_1\phi_1}\ .
\end{equation}
Let us rewrite the projection conditions in equations (\ref{projection}) and (\ref{projcomplx}) as:
\begin{equation}
(i\Gamma_{12})\,\eta=-\eta\ ,~~~(i\Gamma_{\hat 1\hat 2})\,\eta=\eta\ ,~~~i(\Gamma_{12})\,\eta^*=\eta^*\,,~~~i(\Gamma_{\hat 1\hat 2})\,\eta^*=-\eta^*\,.
\end{equation}
Then we have:
\begin{eqnarray}
[(i\Gamma_{12}),{\cal P}]\,\eta&=&\{A,(i\Gamma_{12})\}\eta^*\,,\\ \nonumber
[(i\Gamma_{\hat 1\hat 2}),{\cal P}]\,\eta&=&\{A,(i\Gamma_{\hat 1\hat 2})\}\eta^*\,.
\end{eqnarray}
Therefore we need $\{A,\Gamma_{12}\}=\{A,\Gamma_{\hat 1\hat 2}\}=0$, which implies $\{\gamma_{\theta_1\phi_1},\Gamma_{12}\}=0$ and $\{\gamma_{\theta_1\phi_1},\Gamma_{\hat 1\hat 2}\}=0\,$. The only matrix on the right-hand side of equation (\ref{gammat1p1}) which anticommutes with both $\Gamma_{12}$ and $\Gamma_{\hat 1 \hat 2}$ is $\Gamma_{1\hat 2}$. Therefore we arrive at the equations:
\begin{equation}
c_1=0\ ,~~~c_4=0\ ,~~~c_5=0\ ,~~~c_6=0\ ,~~~c_7=0\ .
\end{equation}
Equation $c_5=0$ implies $\partial_{\theta_1}\phi_2=0$. The system of equations then reduces to:
\begin{eqnarray}
\label{EOMD5}
&&\cos\theta_1+\cos\theta_2(\partial_{\phi_1}\phi_2)=0\ ,\\
\label{derEOMD5}
&&\sin\theta_1+\sin\theta_2(\partial_{\theta_1}\theta_2)(\partial_{\phi_1}\phi_2)=0\ .
\end{eqnarray}
One can see that equation (\ref{derEOMD5}) is just the derivative of equation (\ref{EOMD5}) with respect to $\theta_1$. Therefore equation (\ref{EOMD5}) is the only independent equation of motion. Furthermore, if we solve for $\partial_{\phi_1}\phi_2$ in equation (\ref{EOMD5}), we obtain:
\begin{equation}
\partial_{\phi_1}\phi_2=-\frac{\cos\theta_1}{\cos\theta_2}=p=\rm{const}\ ,
\end{equation}
because the left-hand side there is a function of $\phi_1$, while the right-hand side is a function of $\theta_1$, and hence the only option is both to be constants. The only restriction that we impose on $p$ is to be an integer different from zero.\footnote{In the analogous analysis for the Poincare patch considered in ref. \cite{Arean:2004mm} the authors made a minor error and concluded that $|p|=1$.} Namely,
\begin{equation}
|p|=1\,,2\,,3\,,\dots\ .
\end{equation}
Therefore the profile of the supersymmetric D5--brane is given by:
\begin{equation}
\label{profileD5}
\theta_2(\theta_1)=\arccos\!\left(-\frac{\cos\theta_1}{p}\right)\,,~~~~\phi_2(\phi_1)=p\,\phi_1+{\rm const}\ .
\end{equation}
Let us consider again the operator ${\cal P}$ defined in (\ref{P}). Clearly we have ${\cal P}^2=1$, which implies $AA^*=1$. For the solution (\ref{profileD5}) the matrix $A$ is given by:
\begin{equation}
A=\frac{i}{\sqrt{g_{X^2}}}\,c_2\,e^{i\psi}\,\Gamma_{t\,\rho\,\alpha_2\,\alpha_3}\,\Gamma_{1\hat 2}\ .
\end{equation}
The condition $AA^*=1$ boils down to:
\begin{equation}
\label{condition}
|c_2|^2=g_{X^2}\ .
\end{equation}
One can check that indeed for the submanifold $X^2$ specified by (\ref{profileD5}) equation (\ref{condition}) is satisfied. We have also checked that the solution (\ref{profileD5}) is a solution of the equations of motion obtained from the DBI action of a D5--brane embedding given by the ansatz~(\ref{ansatzD5}).

Finally, let us discuss the number of supersymmetries preserved by the D5--brane probe. Using the solution for $\theta_2$ and $\phi_2$ one can show that:
\begin{equation}
c_2=-\frac{L^2}{6}\frac{p}{|p|}\frac{p^2-\cos^22\theta_1}{\sqrt{p^2-\cos^2\theta_1}}\ .
\end{equation}
However $|p|\geq1$ and hence ${\rm sign}(c_2)=-{\rm sign}(p)$. The ${\cal P}\,\eta=\eta$ is given on-shell by:
\begin{equation}
{\cal P}\,\eta=i\,\mbox{sign}(p)\,e^{i\psi}\,\Gamma_{t\,\rho\,\alpha_2\,\alpha_3}\,\Gamma_{1\hat 2}\,\eta^*=\eta\ .
\end{equation}
Let us define
\begin{equation}
\eta=e^{i\left(\frac{\psi}{2}+\frac{\pi}{4}\right)}\lambda\ .
\end{equation}
Clearly the spinor $\lambda$ is constant and has the same number of independent components as $\eta$. The projection for $\lambda$ is given by:
\begin{equation}
{\rm sign}(p)\,\Gamma_{t\,\rho\,\alpha_2\,\alpha_3}\,\Gamma_{1\hat 2}\,\lambda^*=\lambda\ ,
\end{equation}
where the matrix on the right-hand side is real. If we decompose $\lambda=\lambda_1+i\lambda_2$, where $\lambda_1$ and $\lambda_2$ are real spinors, we arrive at:
\begin{eqnarray}
{\rm sign}(p)\,\Gamma_{t\,\rho\,\alpha_2\,\alpha_3}\,\Gamma_{1\hat 2}\,\lambda_1&=&\lambda_1\ ,\\ \nonumber
{\rm sign}(p)\,\Gamma_{t\,\rho\,\alpha_2\,\alpha_3}\,\Gamma_{1\hat 2}\,\lambda_2&=&-\lambda_2\ .
\end{eqnarray}
One can see that half of the degrees of $\lambda_1$ and half of the degrees of $\lambda_2$ are projected out, and therefore the D5--brane embedding preserves half of the number of supersymmetries of the background, which amounts to four supercharges.

\subsection{Kappa symmetry for a D7--brane.}
For a D7--brane the kappa-symmetry matrix is given by:
\begin{equation}
\Gamma_{\kappa}=-\frac{i}{8!\sqrt{-g}}\,\varepsilon^{\mu_1\dots\mu_8}\gamma_{\mu_1\dots\mu_8}\ .
\end{equation}
We choose the following ansatz for the D7--brane embedding:
\begin{eqnarray}
\xi^{\mu}&=&(t,\,\rho,\,\alpha_1,\,\alpha_2,\,\alpha_3,\,\theta_1,\,\phi_1,\,\phi_2)\ ,\\
\psi&=&\psi(\phi_1,\,\phi_2)\ ,~~~~\theta_2=\theta_2(\rho,\,\theta_1)\ .
\end{eqnarray}
The kappa-symmetry matrix reduces to:
\begin{equation}
\Gamma_{\kappa}=-\frac{i}{\sqrt{-g}}\,\gamma_{t\,\rho\,\alpha_1\,\alpha_2\,\alpha_3\,\theta_1\,\phi_1\,\phi_2}\ ,
\end{equation}
where:
\begin{eqnarray}
\gamma_t&=&L\,\cosh\rho\,\Gamma_t\ ,~~~~~~~~~~~~~\gamma_{\alpha_1}=L\,\sinh\rho\,\Gamma_{\alpha_1}\ ,\\ \nonumber
\gamma_{\alpha_2}&=&L\,\sinh\rho\,\sin\alpha_1\,\Gamma_{\alpha_2}\ ,~~~\,\gamma_{\alpha_3}=L\,\sinh\rho\,\sin{\alpha_1}\,\sin\alpha_2\,\Gamma_{\alpha_3}\ ,\\ \nonumber
\gamma_{\rho}&=&L\,\Gamma_{\rho}+\frac{L}{\sqrt{6}}\,\partial_{\rho}\theta_2\left(\cos\psi\,\Gamma_{\hat 1}+\sin\psi\,\Gamma_{\hat 2}\right)\,,\\ \nonumber
\gamma_{\theta_1}&=&\frac{L}{\sqrt{6}}\,\Gamma_1+\frac{L}{\sqrt{6}}\,\partial_{\theta_1}\theta_2\left(\cos\psi\,\Gamma_{\hat 1}
+\sin\psi\,\Gamma_{\hat 2}\right)\,,\\ \nonumber
\gamma_{\phi_1}&=&\frac{L}{\sqrt{6}}\,\sin\theta_1\,\Gamma_2+\frac{L}{3}\left(\partial_{\phi_1}\psi+\cos\theta_1\right)\Gamma_3\ ,\\ \nonumber
\gamma_{\phi_2}&=&\frac{L}{\sqrt{6}}\,\sin\theta_2\left(\sin\psi\,\Gamma_{\hat 1}-\cos\psi\,\Gamma_{\hat 2}\right)
+\frac{L}{3}\left(\partial_{\phi_2}\psi+\cos\theta_2\right)\Gamma_3\ .
\end{eqnarray}
For the $\Gamma_{\kappa}$-symmetry matrix we obtain:
\begin{equation}
\Gamma_{\kappa}=\frac{i}{{\rm Vol}\,X^3}
\left[-\gamma_*\gamma_{[\theta_1\,\phi_1\,\phi_2]}+\frac{1}{\sqrt{6}}\,\partial_{\rho}\theta_2\,\Gamma_{t\,\alpha_1\,\alpha_2\,\alpha_3}
\left(\cos\psi\,\Gamma_{[\hat 1}+\sin\psi\,\Gamma_{[\hat 2}\right)\gamma_{\theta_1\,\phi_1\,\phi_2]}\right]\,,\label{Gam7interm}
\end{equation}
where $X^3$ is the three-dimensional submanifold of the internal $S^5$ wrapped by the D7--brane. The second term in equation (\ref{Gam7interm}) does not commute with the matrices in the definition of the Killing spinor (\ref{kilspin}). Therefore we need to set $\partial_{\rho}\theta_2=0$. We arrive at:
\begin{equation}
\Gamma_{\kappa}=-\frac{i}{{\rm Vol}\,X^3}\,\gamma_*\gamma_{[\theta_1\,\phi_1\,\phi_2]}\ .
\end{equation}
By using the projections (\ref{projectioneps}) one can show that:
\begin{equation}
\label{gammaS3}
\gamma_{[\theta_1\,\phi_1\,\phi_2]}\,\varepsilon=C_1i\Gamma_{\hat 2}\,\varepsilon+C_2i\Gamma_3\,\varepsilon
+C_3\Gamma_{13\hat 2}\,\varepsilon+C_4i\Gamma_2\,\varepsilon\ ,
\end{equation}
where:
\begin{eqnarray}
C_1&=&-\frac{L^3}{6\sqrt{6}}\sin\theta_1\,\sin\theta_2\,e^{-i\psi}\,,~~
C_2=\frac{L^3}{18}\,[\sin\theta_1(\partial_{\phi_2}\psi+\cos\theta_2)-\sin\theta_2\partial_{\theta_1}\theta_2]\ ,\nonumber\\
C_3&=&-\frac{L^3}{6\sqrt{6}}\sin\theta_2\,(\partial_{\phi_1}\psi+\cos\theta_1)\ ,~~
C_4=-\frac{L^3}{6\sqrt{6}}\sin\theta_1\sin\theta_2\,\partial_{\theta_1}\theta_2\ .
\end{eqnarray}
On the other hand the projection $\Gamma_{\kappa}\,\varepsilon=\varepsilon$ should be compatible with the projections (\ref{projectioneps}). Therefore we need $[\Gamma_{\kappa},\Gamma_{12}]=[\Gamma_{\kappa},\Gamma_{\hat1\hat2}]=0$. However, $\gamma_*$ commutes with $\Gamma_{12}$ and $\Gamma_{\hat1\hat2}$, therefore we need $\gamma_{[\theta_1\,\phi_1\,\phi_2]}$ to commute with $\Gamma_{12}$ and $\Gamma_{\hat1\hat2}$. It is easy to check that only the term $\Gamma_3\,\varepsilon$ in equation (\ref{gammaS3}) satisfies this condition. Therefore we need to set $C_1,\,C_3$ and $C_4$ to zero, resulting in the equations:
\begin{equation}
\sin\theta_1\,\sin\theta_2=0\ ,~~~\sin\theta_2\,(\partial_{\phi_1}\psi+\cos\theta_1)=0\ ,~~~\sin\theta_1\sin\theta_2\,\partial_{\theta_1}\theta_2=0\ ,
\end{equation}
which imply that $\sin\theta_2=0$, hence $\theta_2=0$ or $\pi$. The D7--brane embedding is then described by:
\begin{equation}
\label{solutionD7}
\theta_2=0\,,\,\pi\ ,~~~\psi=n_1\phi_1+n_2\phi_2+{\rm const}\ ,~~~n_1,n_2\in\IZ~{\rm and}~n_2\neq-\cos\theta_2\ ,
\end{equation}
where the restriction $n_2\neq-\cos\theta_2$ is imposed, because at $n_2=-\cos\theta_2$ the $\phi_2$ cycle collapses, and $X^3$ is not a three-dimensional manifold anymore.

Let us analyze the amount of preserved supersymmetry. One can show that the component of the volume form of $X^3$ is given by:
\begin{equation}
{\rm Vol}\,X^3=\frac{L^3}{18}\sin\theta_1\big|\partial_{\phi_2}\psi+\cos\theta_2\big|=|C_2|\ .
\end{equation}
On the other hand, for the allowed values of $n_2$ one can easily check that $C_2={\rm sign}(n_2)|C_2|$. The projection $\Gamma_{\kappa}\,\varepsilon=\varepsilon$ is given by:
\begin{equation}
{\rm sign}(n_2)\Gamma_{t\,\rho\,\alpha_1\,\alpha_2\,\alpha_3\,\hat 3}\,\varepsilon=\varepsilon\ ,
\end{equation}
which can also be written as a projection on the constant spinor $\eta$:
\begin{equation}
{\rm sign}(n_2)\Gamma_{t\,\rho\,\alpha_1\,\alpha_2\,\alpha_3\,\hat 3}\,\eta=\eta\ .
\end{equation}
One can see that half of the degrees of freedom of $\eta$ are projected out, and hence the embedding preserves half of the original supersymmetry of the background, which again amounts to four supersymmetries.

%
%
\section{Meson spectrum of D7--brane embeddings}
In this section we consider particular case of a supersymmetric D7--brane embedding and analyse the spectrum of fluctuations. We will show that the ground state of the spectra is given by the conformal dimension of the dual meson operator. The structure of the spectrum suggests that the Casimir energy dissociates the meson spectrum.
\subsection{General remarks about the D7--brane embedding.}
It is convenient to rewrite the metric of $AdS_5\times T^{1,1}$ in a new radial coordinate $r=\sinh\rho$. The metric in these coordinates is given by:
\begin{eqnarray}
ds^2=-\left(1+\frac{r^2}{R^2}\right)d\tau^2+r^2d\Omega_3^2+\frac{dr^2}{1+\frac{r^2}{R^2}}+R^2ds_{T^{1,1}}^2\ ,
\end{eqnarray}
and the metric on $T^{1,1}$ is given by:
\begin{equation}
ds_{T^{1,1}}^2=\frac{1}{6}\sum\limits_{i=1}^2\left(d\theta_i^2+\sin^2\theta_i\,d\phi_i^2\right)
+\frac{1}{9}\left(d\psi+\sum\limits_{i=1}^2\cos\theta_i\,d\phi_i\right)^2\,.\label{t11metric}
\end{equation}

Let us consider the supersymmetric D7--brane embeddings specified by equation (\ref{solutionD7}). Without loss of generality we can consider embeddings with $\theta_2=0$, and for simplicity we will restrict our considerations to the $n_1=n_2=1$ case. In fact it will be more convenient to redefine $\psi\to\psi+\phi_1+\phi_2+{\rm const}$ and write the metric on $T^{1,1}$ as
\begin{equation}
ds_{T^{1,1}}^2=\frac{1}{6}\sum\limits_{i=1}^2\left(d\theta_i^2+\sin^2\theta_i\,d\phi_i^2\right)
+\frac{1}{9}\left(d\psi+\sum\limits_{i=1}^2(1+\cos\theta_i)\,d\phi_i\right)^2\,.
\end{equation}
In these coordinates the D7--brane embedding is extended along $\tau,\,\Omega_3,\,r,\,\theta_1\,,\phi_1,\,\phi_2$ and is at $\theta_2=0,\psi={\rm const}$. Since the internal manifold wrapped by the D7--brane does not depend on the holographic coordinate $\rho$, this embedding corresponds to the addition of a massless flavour to the dual gauge theory. The induced metric $g_{\alpha\beta}=\partial_{\alpha}X^{\mu}\partial_{\beta}X^{\nu}G_{\mu\nu}$ on the D7--brane's worldvolume is given by:
\begin{eqnarray}
\label{m8metric}
ds_{{\cal M}_8}^2&=&-\left(1+\frac{r^2}{R^2}\right)d\tau^2+r^2d\Omega_3^2+\frac{dr^2}{1+\frac{r^2}{R^2}}+{R^2}ds_{S_q^3}^2\ ,\\
ds_{S_q^3}^2&=&\frac{1}{6}(d\theta_1^2+\sin^2\theta_1\,d\phi_1^2)+\frac{1}{9}(2d\phi_2+(1+\cos\theta_1)\,d\phi_1)^2\ ,
\label{sq3metric}
\end{eqnarray}
where $S_q^3$ stands for a squashed 3-sphere. The DBI action of the probe brane is given by:
\begin{equation}
S=\frac{\mu_7}{g_s}\int d^8\xi\sqrt{-{\rm det}\,g}\ ,
\end{equation}
where:
\begin{eqnarray}
\sqrt{-{\rm det}\,g}&=&{\cal G}(r)\sqrt{|g_{S_{}^3}|}\,\sqrt{|g_{S_q^3}|}\ ,\quad{\cal G}(r)=R^3r^3\,,\label{detgd7}\\
\sqrt{|g_{S_{}^3}|}&=&\sin^2\alpha_1\sin\alpha_2\ ,\quad\sqrt{|g_{S_q^3}|}=\frac{1}{9}\sin\theta_1\ .\nonumber
\end{eqnarray}

\subsection{Meson spectrum.}
To study the meson spectrum of the theory, we consider fluctuations of the transverse scalars:
\begin{equation}
\theta_2=0+(2\pi\alpha')\delta\theta_2\ ,\quad\psi=0+(2\pi\alpha')\delta\psi\ ,
\end{equation}
and expand the DBI action to second order in $\alpha'$. We obtain:
\begin{eqnarray}
\frac{{\cal L}_{\theta_2}^{(2)}}{(2\pi\alpha')^2}&=&\frac{G_{\theta_2\theta_2}}{2}\sqrt{-{\rm det}\,g}\left(g^{\alpha\beta}\,\partial_{\alpha}\delta\theta_2\,\partial_{\beta}\delta\theta_2+\frac{3(1+3\cos\theta_1)}{4(1-\cos\theta_1)\,R^2}\,\delta\theta_2^2\right)\ ,\\
\frac{{\cal L}_{\psi}^{(2)}}{(2\pi\alpha')^2}&=&\frac{\sqrt{-{\rm det}\,g}}{2}\left[G_{\psi\psi}-G_{\psi\phi_i}\,G_{\psi\phi_j}g^{\phi_i\phi_j}\right]
g^{\alpha\beta}\,\partial_{\alpha}\,\delta\psi\,\partial_{\beta}\,\delta\psi\ .\label{Lagr-psi}
\end{eqnarray}
One can see that the scalar excitations do not couple for this choice of coordinates. Let us first analyse the spectrum of fluctuations along $\theta_2$. Furthermore, if one evaluates the coefficient $\left[G_{\psi\psi}-G_{\psi\phi_i}\,G_{\psi\phi_j}g^{\phi_i\phi_j}\right]$ in equation (\ref{Lagr-psi}) at $\theta_2=0$ one obtains a zero. This suggests that for massless (and supersymmetric) embeddings the fluctuations along $\psi$ are not well defined, in a sense fluctuations along the angular variable $\psi$ correspond to rotations with zero radius, which are not well defined. 

One could ratify this by giving a small mass $\mu$ to the quarks and analysing the leading terms in (\ref{Lagr-psi}) in the limit $\mu\to0$. However, the embeddings corresponding to massive quarks have non-trivial radial dependence \cite{Arean:2004mm} and for the theory on a three-sphere they do not correspond to supersymmetric embeddings, making them significantly more difficult to analyse. This is why we focus on the spectrum of supersymmetric embeddings and analyse the spectrum of fluctuations along $\theta_2$.

The equation of motion for $\delta\theta_2$ is given by:
\begin{equation}
\frac{1}{G_{\theta_2\theta_2}\sqrt{-{\rm det}\,g}}\,\,\partial_{\alpha}\left(G_{\theta_2\theta_2}\sqrt{-{\rm det}\,g}\,g^{\alpha\beta}\partial_{\beta}\,\delta\theta_2\right)-\frac{3(1+3\cos\theta_1)}{4(1-\cos\theta_1)\,R^2}\,\delta\theta_2=0\ .
\end{equation}
Next we consider the ansatz for $\delta\theta_2$:
\begin{equation}
\delta\theta_2=\eta(r)e^{i\omega\tau}{\cal Y}^l(S^3)\zeta(S_q^3)\ .
\end{equation}
After splitting $\sqrt{-{\rm det}\,g}$ as in equation (\ref{detgd7}), we obtain:
\begin{eqnarray}
-g^{00}\omega^2+\frac{1}{{\cal G}(r)\,\eta(r)}\,\partial_r\left({\cal G}(r)\,g^{rr}\,\partial_r\,\eta(r)\right)&+&\frac{1}{r^2}\left(\frac{\Delta_3{\cal Y}_l}{{\cal Y}_l}\right)+\label{EOMth}\\
&+&\frac{1}{R^2}\left(\frac{\hat\Delta_3\,\zeta}{\zeta}\right)-\frac{3(1+3\cos\theta_1)}{4(1-\cos\theta_1)\,R^2}=0\ .
\nonumber
\end{eqnarray}
Using that $\Delta_3\,{\cal Y}_l=-l(l+2)\,{\cal Y}_l$ and that the last two terms in equation (\ref{EOMth}) are independent on $r$, we can split variables to obtain:
\begin{eqnarray}
\frac{1}{r^3}\partial_r\left(r^3\left(1+\frac{r^2}{R^2}\right)\,\partial_r\eta(r)\right)
+\left[\frac{\omega^2}{1+\frac{r^2}{R^2}}-\frac{l(l+2)}{r^2}+\frac{\kappa}{R^2}\right]\,\eta(r)&=&0\ ,\label{EOMet}\\
\hat\Delta_3\,\zeta(S_q^3)-\left[\frac{3(1+3\cos\theta_1)}{4(1-\cos\theta_1)}+\kappa\right]\zeta(S_q^3)&=&0\ .\label{EOMzeta}
\end{eqnarray}
Before we quantise the spectrum, let us calculate the conformal dimension of the operators corresponding to $\delta\theta_2$. To this end we solve equation (\ref{EOMet}) near the boundary of the AdS$_5$ space. According to the standard AdS/CFT dictionary the leading mode should behave as $\propto r^{k_1}=r^{\Delta-4+p}$, while the subleading one should scale as $\propto r^{k_2}=r^{-\Delta+p}$ for some constant $p$. Therefore one can calculate the conformal dimension from:
\begin{equation}
\Delta=2+(k_1-k_2)/2\ .\label{dimint}
\end{equation}
The asymptotic form of equation (\ref{EOMet}) for $r\gg R$ is given by:
\begin{equation}
\frac{d^2}{dr^2}\eta(r)+\frac{5}{r}\,\frac{d}{dr}\eta(r)+\frac{\kappa}{r^2}\eta(r)=0\ ,
\end{equation}
with a general solution
\begin{equation}\label{large-r}
\eta(r)=C_1\,r^{-2+\sqrt{4-\kappa}}+C_2\,r^{-2-\sqrt{4-\kappa}}\ .
\end{equation}
After applying equation (\ref{dimint}) we obtain
\begin{equation}
\Delta=2+\sqrt{4-\kappa}\label{conf-dim}
\end{equation}
for the conformal dimensions of the operators corresponding to the excitations of $\delta\theta_2$. Note that the parameter $\kappa$ (related to the AdS mass of the scalar fluctuations) is quantised by solving equation~(\ref{EOMzeta}). However we can still extract non-trivial information about the spectrum of excitations from solving equation (\ref{EOMet}) first. Indeed, one can show that the solution regular at $r=0,\,\infty$ is given by:
\begin{equation}
\eta(r)=\frac{r^{l}}{(1+r^2)^{\frac{\omega R}{2}}}\,_2F_1\left[\frac{1}{2}(2+\sqrt{4-\kappa}+l-{\omega}{R}), \frac{1}{2}(2-\sqrt{4-\kappa}+l-{\omega}{R}),2+l,-r^2\right].\label{regsolet}
\end{equation}
One can check that for this choice of $\eta(r)$ the coefficient $C_1$ (\ref{large-r}) vanishes, thus making the mode normalisable. 
After quantising the first argument in equation~(\ref{regsolet}), we obtain:
\begin{equation}
\omega=\frac{1}{R}(2+\sqrt{4-\kappa}+2n+l)=\frac{1}{R}(\Delta+2n+l)\ ,\quad n,\,l=0,1,2,\dots\ .\label{spectrum-thet2}
\end{equation}

Therefore the ground state of the spectrum of quantum fluctuations is given by the conformal dimension of the corresponding operator in units of the inverse radius of the $S^3$ where the dual conformal theory is defined, which is the same relation that was uncovered in ref.~\cite{Erdmenger:2010zm} for the case of flavoured ${\cal N}=4$ SYM theory. The equidistant structure and the fact that the energies of excitations with the same angular momentum quantum number $l$ differ, by an even number $2n$ times $1/R$, can be understood as the result of a summation of angular momentum of the fundamental fields composing the bound state (the meson). This supports the interpretation that the mesons are dissociated at zero bare mass due to the effect of the finite Casimir energy of the theory on a three-sphere. In this presentation the conformal dimension $\Delta$ in equation (\ref{spectrum-thet2}) is related to the total zero point energy of the fields building the meson state. 

Following ref.~\cite{Erdmenger:2010zm} we show in details how the equidistant structure of the spectrum (\ref{spectrum-thet2}) arises from summation of angular momentum. The operators corresponding to the fluctuations of the D7--brane are of the form \cite{Levi:2005hh}:
\begin{equation}
\tilde q \,\mathcal{O} q \propto \tilde q\,  (AB) (AB)\dots (AB)\,q\ ,
\end{equation}
where $q$ and $\tilde q$ are fundamental fields of appropriate dimension ($3/2$ in our case) transforming in the colour $(\bf N,1)_c$ and $(\bf\bar N,1)_c$ correspondingly, while the operator $\mathcal{O}\propto (AB) (AB)\dots (AB)$ is composed of the bi-fundamental fields $A_\alpha$ and $B_\alpha$ ($\alpha=1,2$) having dimensions $3/4$ and transforming in the colour $(\bf N,\bar N)_c$ and $(\bf \bar N,N)_c$ correspondingly. The pair $(AB)$ is constructed by summing over the fundamental index of $B$ and the anti-fundamental index of $A$ and thus transforms in the $(\bf N,\bar N)_c$ making the meson operator real. If we denote by $E_q^0$ and $E_{\mathcal{O}}^0$ the zero point energies of $q$ and $\mathcal{O}$ and use the fact that energy due to the angular momentum along the $S^3$ is of order $\sim l/R$, we can expand:
\begin{eqnarray}
\mathcal{O}(t,S^3)&=&\frac{1}{\sqrt{2E_{\mathcal{O}}^0}}\left(e^{i\,E_{\mathcal{O}}\,t}X_0+e^{-i\,E_{\mathcal{O}}\,t}X_0^{\dagger} \right)\ , \\
\tilde q(t,S^3)&=&\sum\limits_{l_1=0}^{\infty}\sum\limits_{I_1=0}^{(l_1+1)^2}\frac{\left(e^{i(E_q^0 R+l_1)\frac{t}{R}}a_{l_1I_1}\bar {\cal Y}^{ l_1I_1}(S^3)+e^{-i(E_q^0R+ l_1)\frac{t}{R_3}}
{b}^{\dagger}_{l_1I_1}{\cal Y}^{ l_1I_1}( S^3)\right)}{\sqrt{2(E_q^0+\frac{l_1}{R})}}\ ,\nonumber\\
q(t,S^3)&=&\sum\limits_{l_2=0}^{\infty}\sum\limits_{I_2=0}^{(l_2+1)^2}\frac{\left(e^{i(E_q^0 R+l_2)\frac{t}{R}}b_{l_2I_2}\bar {\cal Y}^{ l_2I_2}( S^3)+e^{-i(E_q^0R+ l_2)\frac{t}{R_3}}
{a}^{\dagger}_{l_2I_2}{\cal Y}^{ l_2I_2}( S^3)\right)}{\sqrt{2(E_q^0+\frac{l_2}{R})}}\ ,\nonumber
\end{eqnarray}
where ${\cal Y}^{lI}(S^3)$ are the scalar spherical harmonics on $S^3$. Also $X_0$ transforms  in the $(\bf N,\bar N)_c$, $a_{l_1I_1}$ transforms in the $(\bf\bar N,1)_c$ and $b_{l_2I_2}$ transforms in the $(\bf N,1)_c$. Note that we have suppressed the angular momentum along $S^3$ of the bi-fundamental operator $\mathcal{O}$, reflecting the fact that the geometry also does not have angular momentum along the three-sphere. Now the meson state can be constructed by acting with $X_0^{\dagger},b^{\dagger}_{l_1,I_1},a^{\dagger}_{l_2,I_3}$ on the vacuum state defined by:
\begin{equation}
a_{l_1I_1}|0\rangle=b_{l_2I_2}|0\rangle={X_0}_{ij}|0\rangle=0 \ .
\end{equation}
The meson state is then given by:
\begin{equation}
\tilde q\, X_0\, q \,|0\rangle=\frac{1}{\sqrt{2 E_\mathcal{O}^0}}\sum_{ l_1,l_2=0}^{\infty}\sum_{I_1=0}^{(l_1+1)^2}\sum_{I_2=0}^{( l_2+1)^2}e^{-i(\Delta+l_1+l_2)\frac{t}{R}}\frac{{\cal Y}^{ l_2I_2}(\tilde S^3){\cal Y}^{ l_1I_1}(\tilde S^3)}{\sqrt{(E_q^0+\frac{l_1}{R})(E_q^0+\frac{l_2}{R})}}a^{\dagger}_{ l_2I_2}{X_0}^{\dagger}b^{\dagger}_{ l_1I_1}|0\rangle \ ,\label{expsphr}
\end{equation}
where we have substituted $2E_q^0+E_\mathcal{O}^0=\Delta/R$. Note that the state (\ref{expsphr}) is a superposition of states with
definite energy $E_{l,J}=(\Delta+J)/R$, where $J=l_1+l_2$. Our next step is to expand the product of the spherical harmonics in (\ref{expsphr}) in a Laplace series:
\begin{equation}
{\cal Y}^{ l_1I_1}(S^3){\cal Y}^{ l_2I_2}(
S^3)=\sum_{l=0}^{\infty}\sum_{I=0}^{( l+1)^2}{\bf
  C}^{lI}_{ l_1I_1,l_2I_2}{\cal Y}^{ l
  I}(S^3)\,  . \label{Lapl}
\end{equation}
The coefficients ${\bf C}^{lI}_{ l_1I_1,l_2I_2}$
are non-zero only for $| l_1- l_2 | \leq  l\leq  l_1+ l_2$ (addition of angular momentum) and
$(-1)^{l}=(-1)^{ l_1+l_2}$(conservation of the
antipodal map eigenvalue)~\cite{Hamada:2003jc}. Therefore the Laplace series in
(\ref{Lapl}) terminates at $J=l_1+ l_2$ and we can write $J=2n+l$ ,where $n$ is an integer non-negative number. This implies that a state with a given energy $E_{Jl}$ can be expanded as:
\begin{equation}
|E_{Jl}\rangle=\frac{1}{\sqrt{2 E_\mathcal{O}^0}}e^{-iE_{Jl}t}\sum_{2n+ l=J}\sum_{I=0}^{( l+1)^2}{\cal Y}^{ l I}(S^3)C_{n l I}^{\dagger}|0\rangle\ ,
\end{equation}
where $C_{n l I}^{\dagger}$ is defined by:
\begin{equation}
C_{n  l I}^{\dagger}\equiv\sum_{l_1+ l_2=2n+l}\sum_{I_1=0}^{( l_1+1)^2}\sum_{I_2=0}^{(l_2+1)^2}\frac{{\bf C}^{ lI}_{ l_1I_1, l_2I_2}}{\sqrt{(E_q^0+\frac{l_1}{R})(E_q^0+\frac{l_2}{R})}}a^{\dagger}_{ l_2I_2}{X_0}^{\dagger}b^{\dagger}_{ l_1I_1}\ .
\end{equation}
Now the meson state (\ref{expsphr}) can be written as:
\begin{equation}
\tilde q\, X_0\, q \,|0\rangle=\frac{1}{\sqrt{2 E_\mathcal{O}^0}}\sum_{n, l=0}^{\infty}\sum_{I=0}^{( l+1)^2}e^{-i(\Delta+l+2n+ l)\frac{t}{R}}{\cal Y}^{ l I}(\tilde S^3)C_{n l I}^{\dagger}|0\rangle\ .\label{collectspect}
\end{equation}
Therefore we obtain the same spectra as in equation (\ref{spectrum-thet2}). This suggest that the conformal dimension of the meson operator is a simple sum of the conformal dimensions of the fundamental operators $q$, $\tilde q$ and $\mathcal{O}$. Note that in general the conformal dimension of $\mathcal{O}$ cannot be obtained as a sum of the engineering dimensions of the bi-fundamental fields $A_\alpha,B_\alpha$ due to contribution from the anomalous dimensions of these operators. Indeed as we are going to show below only the lowest possible conformal dimension can be obtained as a sum of the engineering dimensions of its constituent fields. 

To calculate the spectrum of conformal dimensions we need to quantise the parameter $\kappa$ in equation (\ref{conf-dim}). To this end we substitute the ansatz
\begin{equation}
\zeta(\theta_1,\phi_1,\phi_2)=\hat\zeta(\theta_1)e^{im_1\phi_1}e^{im_2\phi_2}
\end{equation}
in equation (\ref{EOMzeta}) and write down explicitly the equation of motion for $\hat\zeta(\theta_1)$:
\begin{equation}
\frac{1}{\sin\theta_1}\partial_{\theta_1}\left(\sin\theta_1\partial_{\theta_1}\hat\zeta\right)-\left[\frac{m_1^2}{\sin^2\theta_1}
+\frac{8m_1\,m_2-1-5m_2^2+(m_2^2-3)\cos\theta_1}{8(\cos\theta_1-1)}+\frac{\kappa}{6}\right]\hat\zeta=0\ .
\end{equation}
Next we define $2x=1-\cos\theta_1$ and bring the equation of motion for $\zeta(x)$ to the standard form of a hypergeometric equation. One can show that a solution regular at $x=0$ is given by:
\begin{gather}
\zeta(x)=x^{\frac{\sqrt{1+(m_1-m_2)^2}}{2}}(1-x)^{\frac{m_1}{2}}\,_2F_1\!\left[\frac{1+m_1+\sqrt{1+(m_1-m_2)^2}+\sqrt{\frac{15}{6}-\frac{m_2^2}{2}-\frac{2\kappa}{3}}}{2},\right.\\
\left.\frac{1+m_1+\sqrt{1+(m_1-m_2)^2}-\sqrt{\frac{15}{6}-\frac{m_2^2}{2}-\frac{2\kappa}{3}}}{2},1+\sqrt{1+(m_1-m_2)^2}\,,\,x\right].\nonumber
\end{gather}
To truncate the expansion of the hypergeometric function we quantise its first argument. The resulting spectrum for $\kappa$ is:
\begin{gather}
\kappa=-3(1+2m+{m_1})\sqrt{1+(m_1-m_2)^2}+\frac{3}{4}\left[1-8m(1+m+m_1)\,-\nonumber\right.\\
\left.-\,4m_1(1+m_1-m_2)-3m_2^2\right]\ ,\quad\text{where}\quad m,m_1,m_2\,=0,1,2,\dots\ .
\end{gather}
One can see that in general $\kappa$ is an irrational number. Furthermore, for $m_1\neq m_2$ it is always irrational. However for $m=m_1=m_2=0$ one has $\kappa=-\frac{9}{4}$. The corresponding conformal dimension is $\Delta=9/2$. This suggests that the gauge invariant operator is of the form $\tilde{q} (AB)(AB)q$.
\section{Mesons from probe D5--brane}

\subsection{General setup for D5--brane embedding.}
In this section we focus on the spectra of supersymmetric D5--brane embeddings corresponding to fundamental defect theory living on a maximal $S^2$ inside the $S^3$. We write down the metric of $AdS_5\times T^{1,1}$ as:
\begin{eqnarray}
ds^2=-\left(1+\frac{r^2}{R^2}\right)d\tau^2+r^2(d\alpha^2+\sin^2\alpha\,d\Omega_2^2)+\frac{dr^2}{1+\frac{r^2}{R^2}}+R^2ds_{T^{1,1}}^2\ ,
\end{eqnarray}
where the metric on $T^{1,1}$ is the usual one \eqref{t11metric}, with ranges of angles: $0<\theta_i<\pi$, $0<\phi_i<2\pi$ and $0<\psi<4\pi$. The profile of a supersymmetric embedding is given by equation (\ref{profileD5}). We will consider the particular case when the probe D5--brane is extended along $\tau,\,\Omega_2,\,r,\,\theta_1\,,\phi_1$ and with $\alpha=\pi/2,\,\theta_2=\theta_1,\,\phi_2=2\pi-\phi_1,\,\psi={\rm const}$, which corresponds to the addition of a massless flavour to the dual gauge theory. 

To calculate the spectrum of fluctuations it is more convenient to consider the following parametrisation of $T^{1,1}$:
\begin{equation}
\theta_{\pm}=\frac{1}{2}(\theta_1\pm\theta_2);\ ,~~\phi_{\pm}=\frac{1}{2}(\phi_1\pm\phi_2); \ .
\end{equation}
In the new coordinates the metric of $T^{1,1}$ is given by:
\begin{eqnarray}
ds_{T^{1,1}}&=&\frac{1}{3}(d\thetp^2+d\thetm^2)+\frac{1}{6}(1-{\cos2\thetp}\,{\cos2\thetm})(d\phip^2+d\phim^2)+\frac{1}{3}\sin2\thetp+\,\sin2\thetm d\phip d\phim \nonumber\\
&+&\frac{1}{9}(d\psi+2\cos\thetp\cos\thetm d\phip -2\sin\thetp\sin\thetm d\phim)^2\ .
\end{eqnarray}
The classical embedding of the D5--brane is given by $\thetm=0,~\phip=\pi,~\psi=\psi_0={\rm const}$ and the induced metric $g_{\alpha\beta}=\partial_{\alpha}X^{\mu}\partial_{\beta}X^{\nu}G_{\mu\nu}$ on the D5--brane's worldvolume is given by:
\begin{eqnarray}
ds_{{\cal M}_6}^2&=&-\left(1+\frac{r^2}{R^2}\right)d\tau^2+r^2d\Omega_2^2+\frac{dr^2}{1+\frac{r^2}{R^2}}+R^2ds_{S^2}^2\ ,\\
ds_{S^2}^2&=&\frac{1}{3}\left(d\theta^2+\sin^2\theta\, d\phim^2\right)\ ,
\end{eqnarray}
where we introduced $\theta=\thetp$. The DBI action of the probe brane is given by:
\begin{equation}
S=\frac{\mu_5}{g_s}\int d^6\xi\sqrt{-{\rm det}\,g}\ ,
\end{equation}
where:
\begin{eqnarray}
\sqrt{-{\rm det}\,g}&=&{\cal G}(r)\sqrt{|g_{S_{}^2}|}\,\sqrt{|g_{S^2}|}\ ,\quad{\cal G}(r)=R^2r^2\ ,\label{detgd5}\\
\sqrt{|g_{S_{}^2}|}&=&\sin\beta\ ,\quad\sqrt{|g_{S^2}|}=\frac{1}{3}\sin\theta\ .\nonumber
\end{eqnarray}

\subsection{Meson spectrum of D5--brane.}
To study the meson spectrum of the theory, we consider the fluctuations:
\begin{equation}
\thetm=0+(2\pi\alpha')\delta\xi;\ ,~~~\phip=0+(2\pi\alpha')\delta\phip;\ ,~~~\psi=\psi_0+(2\pi\alpha')\delta\psi; \ ,
\end{equation}
and expand the DBI action to second order in $\alpha'$. We obtain:
\begin{eqnarray}
\frac{{\cal L}_{\xi\xi}^{(2)}}{(2\pi\alpha')^2}&\propto&\frac{1}{2}\sqrt{-{\rm det}\,g}\left[G_{\thetm\thetm}\,g^{\alpha\beta}\,\partial_{\alpha}\delta\xi\,\partial_{\beta}\delta\xi
+\left(\frac{1}{3}+\cot^2\theta\right)\delta\xi^2\right]\ , \label{eqntheta}\\
\frac{{\cal L}_{\phip\phip}^{(2)}}{(2\pi\alpha')^2}&\propto&\frac{1}{2}\sqrt{-{\rm det}\,g}\,G_{\phip\phip}\,g^{\alpha\beta}\,\partial_{\alpha}\delta\phip\,\partial_{\beta}\delta\phip\ ,\\
\frac{{\cal L}_{\psi\psi}^{(2)}}{(2\pi\alpha')^2}&\propto&\frac{1}{2}\sqrt{-{\rm det}\,g}\,G_{\psi\psi}\,g^{\alpha\beta}\partial_{\alpha}\delta\psi\,\partial_{\beta}\delta\psi\ ,\\
\frac{{\cal L}_{\psi\phip}^{(2)}}{(2\pi\alpha')^2}&\propto&\sqrt{-{\rm det}\,g}\,G_{\phip\psi}\,g^{\alpha\beta}\,\partial_{\alpha}\delta\phip\,\partial_{\beta}\delta\psi\ ,\\
\frac{{\cal L}_{\xi\phip}^{(2)}}{(2\pi\alpha')^2}&\propto&\sqrt{-{\rm det}\,g}\,\frac{2}{3}\cot\thetp\,\partial_{\phim}\delta\phip\,\delta\xi\ , \label{coupled1}\\ 
\frac{{\cal L}_{\xi\psi}^{(2)}}{(2\pi\alpha')^2}&\propto&-\sqrt{-{\rm det}\,g}\,\frac{2}{3}\csc\thetp\,\partial_{\phim}\delta\psi\,\delta\xi\ .  \label{coupled2}
\end{eqnarray}
From equations (\ref{coupled1}) and (\ref{coupled2}) one can see that as long as the fluctuations have momentum along $\phim$, $\delta\xi$ couples to $\delta\phip$ and $\delta\psi$. To decouple equation (\ref{eqntheta}) we consider the following ansatz:
\begin{equation}
\delta\xi=\eta(r)e^{i\omega\tau}{\cal Y}^l(S^2)\zeta(\theta)\ .
\end{equation}
After splitting $\sqrt{-{\rm det}\,g}$ as in equation (\ref{detgd5}), and using that $\Delta_2\,{\cal Y}_l=-l(l+1)\,{\cal Y}_l$, we can split the variables to obtain:
\begin{eqnarray}
\frac{1}{r^2}\,\partial_r\!\left(r^2\left(1+\frac{r^2}{R^2}\right)\,\partial_r\eta(r)\right)
+\left[\frac{\omega^2}{1+\frac{r^2}{R^2}}-\frac{l(l+1)}{r^2}+\frac{\kappa}{R^2}\right]\eta(r)&=&0\ ,\label{EOMetxi}\\
\frac{1}{\sin\theta}\partial_{\theta}\left(\sin\theta\,\partial_{\theta}\zeta(\theta)\right)-\left[\cot^2\theta+\frac13+\frac{\kappa}{3}\right]\zeta(\theta)&=&0\ .\label{EOMzetaxi}
\end{eqnarray}
In order to calculate the conformal dimension of the operators corresponding to $\delta\xi$ we solve equation (\ref{EOMetxi}) near the boundary of AdS$_5$. The leading mode should behave as $\propto r^{k_1}=r^{\Delta-3+p}$, while the subleading one should scale as $\propto r^{k_2}=r^{-\Delta+p}$ for some constant $p$. Therefore one can calculate the conformal dimension from:
\begin{equation}
\Delta=\frac{3}{2}+\frac{k_1-k_2}{2}\ . \label{conf3D}
\end{equation}
The asymptotic form of equation (\ref{EOMetxi}) for $r\gg R$ is given by:
\begin{equation}
\frac{d^2}{dr^2}\eta(r)+\frac{4}{r}\,\frac{d}{dr}\eta(r)+\frac{\kappa}{r^2}\eta(r)=0\ ,
\end{equation}
with a general solution:
\begin{equation}
\eta(r)=C_1\,r^{-3/2+\sqrt{9/4-\kappa}}+C_2\,r^{-3/2-\sqrt{9/4-\kappa}}\ .
\end{equation}
After applying equation (\ref{conf3D}) we obtain
\begin{equation}
\Delta=\frac{1}{2}\left(3+\sqrt{9-4\kappa}\,\right)
\end{equation}
for the conformal dimensions of the operators corresponding to the excitations of $\delta\xi$. Note that the parameter $\kappa$ (related to the AdS mass of the scalar fluctuations) is quantised by solving equation~(\ref{EOMzetaxi}). However we can still extract non-trivial information about the spectrum of excitations from solving equation (\ref{EOMetxi}) first. Indeed, one can show that the solution regular at $r=0,\,\infty$ is given by:
\begin{eqnarray}\nonumber
&&\eta(r)=\frac{r^{l}}{(R^2+r^2)^{\frac{R\,\omega}{2}}}\,_2F_1\left[\frac{1}{4}(3+2l-\sqrt{9-4\kappa}-2R\omega),\frac{1}{4}(3+2l+\sqrt{9-4\kappa}-2R\omega),\right. \\
&&~~\left. \frac32+l,-\frac {r^2}{L^2}\right]\ .\label{regsoletxi}
\end{eqnarray}
The second argument in equation~(\ref{regsoletxi}) is greater than the first one and after quantizing to truncate the series we obtain:
\begin{equation}
\omega=\frac{1}{R}\left(\frac{3}{2}+\frac{1}{2}\sqrt{9-4\kappa}+2n+l\right)=\frac{1}{R}\big(\Delta+2n+l)\ ,\quad n,l=0,1,2,\dots\ .\label{spectrum-xi}
\end{equation}
Therefore the ground state of the spectrum of quantum fluctuations is given by the conformal dimension of the corresponding operator in units of the inverse radius of the $S^2$ sphere where the dual conformal theory is defined. Furthermore, we observe the same equidistant structure of the spectrum as in the case of D7--brane analysed in the previous section. In fact, one can easily generalise the discussion from the previous section to the case of a defect field theory living on a maximal two-sphere inside the three-sphere. Although we don't know explicitly the composition of the meson operators, they should still have the structure of a sandwich of two fundamental fields $\tilde q$ and $q$ and a bi-fundamental operator $\mathcal{O}$: $\tilde q\,\mathcal{O}\,q$.  The only difference to the case of a D7--brane is that instead of expanding the operators in spherical harmonics on a three-sphere ${\cal Y}^{lI}(S^3)$ one has to expand in the standard spherical harmonics on a two-sphere ${\cal Y}^{lm}(S^2)$ and use the expansion:
\begin{equation}
{\cal Y}^{l_1m_1}(S^2){\cal Y}^{l_2m_2}(S^2)=\sum_{ l=| l_1- l_2|}^{ l_1+
  l_2}\sum_{m=- l}^{l}{\bf C}^{l m}_{
  l_1m_1, l_2m_2}{\cal Y}^{l m}( S^2)\,  . \label{CL-GR-SU2}
\end{equation}
to arrive at equation (\ref{spectrum-xi}).

Let us now quantise the parameter $\kappa$ and the spectrum of the corresponding conformal dimensions. To this end we define $2x=1-\cos\theta$ and bring the equation of motion for $\zeta(x)$ to the standard form of a hypergeometric equation. One can show that a solution regular at $x=0$ is given by:
\begin{gather}
\zeta(x)=[x(1-x)]^{\frac{1}{2}}\,_2F_1\!\left[\frac32-\frac{\sqrt{33-12\kappa}}{6}\, ,\frac32+\frac{\sqrt{33-12\kappa}}{6}\,,2,\,x\right].\nonumber
\end{gather}
To truncate the expansion of the hypergeometric function we quantise its first argument. The resulting spectrum for $\kappa$ is:
\begin{equation}
\kappa=-4-9m-3m^2\ ,\quad\text{where}\quad m\,=0,1,2,\dots\ .
\end{equation}
Therefore for the conformal dimension we obtain:
\begin{equation}
\Delta=\frac{1}{2}(3 + \sqrt{25 + 12 m (3 + m)})\ .
\end{equation}
Note that in general the conformal dimension is an irrational number. However for $m=0$ we have that $\Delta=4.$ It would be interesting to construct explicitly the operator corresponding to this conformal dimension. 


\section{Conclusion}
In this paper we introduced fundamental flavour to the Klebanov--Witten model on a three-sphere. Using a holographic description we considered both the cases of probe D7-- and D5--branes corresponding to $3+1$ and $2+1$ dimensional flavours.

In section 2 using kappa-symmetry we classified the possible space-filling supersymmetric embeddings. Similarly to the case of flavoured ${\cal N}=4$ SYM on a three-sphere, we concluded that supersymmetric embeddings are possible only at zero bare quark mass. This reflects the properties of the supersymmetry algebra on a three-sphere. In the case of D5--branes the supersymmetric embeddings correspond to fundamental defect field theories living on a maximal $S^2$.

In section 3 we analysed the meson spectrum of the flavoured holographic gauge theories dual to the supersymmetric set-ups analysed in section 2. We showed that the meson spectrum has equidistant structure with ground state given by the conformal dimension of the meson operator divided by the radius of $S^3$. We showed that this equidistant structure is the result of the addition of the angular momentum of the fundamental fields composing the meson operator. This supports the interpretation that for zero bare mass the meson states are dissociated to the their constituent fields due to the Casimir energy of the theory. For the case of D7--branes we showed that the lowest conformal dimension is $\Delta=9/2$, which is consistent with the identification of the dual operators suggested in ref.~\cite{Levi:2005hh}. For the case of D5--branes we showed that the lowest conformal dimension is $\Delta=4$. However, we did not identify the corresponding dual operator. We would like to return to this interesting task in a future work.

Finally, it is worth noting that, while we were successful in explaining the equidistant structure of the meson spectra using the same approach as in ref.~\cite{Erdmenger:2010zm} for the case of flavoured ${\cal N}=4$ theory, we could not derive the zero point energies of the constituent fields of the mesons from field theory calculations only. This is in contrast to the flavoured ${\cal N}=4$ theory on a three-sphere, where the free theory could predict the conformal dimensions and the zero point energies. This discrepancy is not a surprise, because the flavoured ${\cal N}=4$ theory has an ${\cal N}=2$ supersymmetry, which suggests the existence of a non-renormalisation theorem. On the other side, in our case the flavours break the ${\cal N}=1$ supersymmetry of the Klebanov--Witten model down to ${\cal N}=1/2$ supersymmetry, which is insufficient for a non-renormalisation theorem. This also suggests that the equidistant structure of the spectra reported above is entirely due to the fact that we have a conformal field theory on a three-sphere (the flavours are massless) and is not related to the amount of supersymmetry.

\section{Acknowledgments}
The authors would like to thank N. Bobev for collaborating at the early stages of this project. The work of V.F. was funded by an INSPIRE IRCSET-Marie Curie International Mobility Fellowship. R.R. acknowledges the support of the Austrian Science Fund (FWF) project I 1030-N16.

\appendix

\section{The parametrization of $T^{1,1}$}
The conifold defined in $\mathbb{C}^4\times\mathbb{C}^4$ by the equation
\begin{equation}
z_1\,z_2=z_3\,z_4
\end{equation}
can be parameterized as follows:
\begin{align}
&z_1=r^{3/2}e^{i/2(\psi-\phi_1-\phi_2)}\sin\frac{\theta_1}{2}\sin\frac{\theta_2}{2}\ ,\qquad
z_2=r^{3/2}e^{i/2(\psi+\phi_1+\phi_2)}\cos\frac{\theta_1}{2}\cos\frac{\theta_2}{2}\ ,\nonumber\\
&z_3=r^{3/2}e^{i/2(\psi+\phi_1-\phi_2)}\cos\frac{\theta_1}{2}\sin\frac{\theta_2}{2}\ ,\qquad z_4=r^{3/2}e^{i/2(\psi-\phi_1+\phi_2)}\sin\frac{\theta_1}{2}\cos\frac{\theta_2}{2}\ .
\label{z-param}
\end{align}
The conifold can be defined also as:
\begin{equation}
\sum\limits_{i=1}^4\omega_i^2=0\ ,
\end{equation}
where $\omega_i$ are related to $z_i$ through:
\begin{align}
&z_1=\omega_1+i\omega_2\ ,\qquad z_2=\omega_1-i\omega_2\ ,\nonumber\\
&z_3=-\omega_3+i\omega_2\ ,\qquad z_4=\omega_3+i\omega_4\ .
\end{align}
Another helpful parametrization is by making use of homogeneous coordinates $(A_a,B_b)$, which are related to $z_i$ by:
\begin{align}
&z_1=A_1B_1\ ,\qquad z_2=A_2B_2\ ,\nonumber\\
&z_3=A_1B_2\ ,\qquad z_4=A_2B_1\ .
\end{align}
The set $(A_a,B_b)$ transforms under $(N,\bar{N})$ and $(\bar{N},N)$ of the gauge group, correspondingly.

\section{Gauge theory side}
The dual conformal field theory is known as the Klebanov--Witten model \cite{Klebanov:1998hh} and is constructed considering a stack of D3--branes placed at the tip of a conifold. It is $\N=1$ supersymmetric $U(N)\times U(N)$ gauge theory with two chiral multiplets $A_i$ in $(N,\overline{N})$ and another two, usually denoted by $B_i$, in $(\overline{N},N)$. The angular part of the conifold is $T^{1,1}$, and its isometries determine the global symmetries of the gauge theory. Being a $U(1)$ bundle over $S^2\times S^2$, this theory obviously has $SU(2)\times SU(2)$ global symmetry which acts separately on the doublets $A_i,B_i$, and also a non-anomalous $U(1)$ R--symmetry.

The most general superpotential which respects the $SU(2)\times SU(2)\times U(1)_R$ symmetry is a quartic superpotential of the form:
\begin{equation}\label{suppot}
W=\frac{g}{2}\,\epsilon^{ij}\epsilon^{kl}\,\mathrm{Tr}A_iB_kA_jB_l\ .
\end{equation}
Note that there is also a $\mathbf{Z}_2$ symmetry. In the geometric picture, i.e. on the conifold, it acts as a reflection, and from the gauge theory point of view it exchanges the two pairs $A_i$ and $B_j$.

The AdS/CFT correspondence suggests that the anomalous dimensions of gauge theory operators are encoded in the dispersion relations in the dual string theory. The latter are expressible in terms of the following three angular momenta:
\begin{equation}
J_A\equiv P_{\phi_1}\ ,\quad J_B\equiv P_{\phi_2}\ ,\quad J_R\equiv P_\psi\ .
\end{equation}

Our mesons are composite operators constructed out of $A_i$ and $B_j$. Then, it is natural to suggest a correspondence between the quantum numbers in the string theory and those of the dual operators. As it was shown in \cite{Klebanov:1998hh}, the strings moving in $T^{1,1}$ are dual to pure scalar operators, i.e. they do not contain fermions, covariant derivatives or gauge field strengths. One can construct scalars by making use of the fact that they are in the bi-fundamental representation. Therefore, the gauge singlets have the form:
\begin{equation}
\mathrm{Tr}\Big(A\:B\cdots A\:\bar{A}\cdots\bar{B}\:B\cdots\bar{B}\:\bar{A}\cdots\Big)\ .
\end{equation}
This form of the operators suggests the correspondence:
\begin{align}
J_A&\longleftrightarrow\frac12\Big[\#(A_1)-\#(A_2)+\#(\overline{A}_2)-\#(\overline{A}_1)\Big]\ ,\\
J_B&\longleftrightarrow\frac12\Big[\#(B_1)-\#(B_2)+\#(\overline{B}_2)-\#(\overline{B}_1)\Big]\ ,\\
J_R&\longleftrightarrow\frac14\Big[\#(A_i)+\#(B_i)-\#(\overline{A}_i)-\#(\overline{B}_i)\Big]\ ,
\end{align}
where $\#(A_1)$ is the number of $A_1$'s under the trace of the dual composite operator, etc.

We note that there exists an inequality between the bare dimension and the R--charge, which is quite natural when written in terms of string variables:
\begin{equation}
E\geq3|J_R|\ .
\label{bound}
\end{equation}
On the gauge theory side it comes from the unitarity bound of the $\N=1$ superconformal algebra. When the bound is saturated, the primary fields close a chiral ring. The complete dictionary between conserved charges in the string theory and the dual gauge theory operators remains an open problem.

\subsection{Adding flavours}

To add $3+1$ dimensional massive flavours one has to add \cite{Levi:2005hh} the following terms to the super potential (\ref{suppot}):
\begin{eqnarray}
W_f&=&W_{flavours}+W_{masses}\\
W_{flavours}&=&h\,\tilde q\,A_1\,Q+g\,q\,B_1\,\tilde Q\ ,~~~~~W_{masses}=\mu_1\,q\,\tilde q+\mu_2\,Q\,\tilde Q\ ,
\end{eqnarray}
where the constants $\mu_1,\mu_2$ are related to the bare masses of the flavour fields $q,Q$. Assuming that one of the masses $\mu_i$ is larger than the other and integrating out the associated flavours, one can obtain a quartic superpotential of the form \cite{Levi:2005hh}:
\begin{equation}
W_f=q\,(A_1\,B_1-\mu)\,\tilde q\ .
\end{equation}
Where the fundamental index of $A_1$ is contracted with the anti-fundamental index of $B_1$ and the parameter $\mu$ is related to the bare quark mass.

\end{document}